# Analysis of Transferred Pre-Trained Deep Convolution Neural Networks in Breast Masses Recognition

Qusay Shihab Hamad[1,3], Hussein Samma[2], Shahrel Azmin Suandi[1*]

[1,2]School of Electrical and Electronic Engineering, Engineering Campus, Universiti Sains Malaysia, 14300 Nibong Tebal, Penang, Malaysia
[2]SDAIA-KFUPM Joint Research Center for Artificial Intelligence (JRC-AI) King Fahd University of Petroleum and Minerals, Dhahran, Saudi Arabia
[3]University of Information Technology and Communications (UOITC), Baghdad, Iraq

*Corresponding author email:* shahrel@usm.my

**Abstract**

Breast cancer detection based on pre-trained convolution neural network (CNN) has gained much interest among other conventional computer-based systems. In the past few years, CNN technology has been the most promising way to find cancer in mammogram scans. In this paper, the effect of layer freezing in a pre-trained CNN is investigated for breast cancer detection by classifying mammogram images as benign or malignant. Different VGG19 scenarios have been examined based on the number of convolution layer blocks that have been frozen. There are a total of six scenarios in this study. The primary benefits of this research are twofold: it improves the model's ability to detect breast cancer cases and it reduces the training time of VGG19 by freezing certain layers. To evaluate the performance of these scenarios, 1693 microbiological images of benign and malignant breast cancers were utilized. According to the reported results, the best recognition rate was obtained from a frozen first block of VGG19 with a sensitivity of 95.64 %, while the training of the entire VGG19 yielded 94.48%.

*Keywords:* Breast cancer, Deep Learning, CNN, Medical Imaging, Transfer learning.

## 1. Introduction

According to the World Health Organization (WHO) (*WHO Position Paper on Mammography Screening*, n.d.), breast cancer is leads to the killing of over half a million women annually. Approximately 53% of these instances originate in developing countries, which account for 82% of the world's population. Breast cancer is the largest cause of cancer-related deaths among women in undeveloped countries and the second highest cause (after lung cancer) among women in developed countries (Gardezi et al., 2019). In areas with limited medical resources, the majority of breast cancer patients are identified at a late stage; their five-year survival rates range from 10 to 40%. On the other hand, in developed countries where early diagnosis and treatment are available, the five-year survival rate is over 80% (*WHO Position Paper on Mammography Screening*, n.d.). This demonstrates the significance of a fast and precise diagnosis of this type of cancer to the patient's survival chances. Since early diagnosis can have a significant impact on the patient's recovery and treatment, many attempts have been made in recent years to improve early diagnosis. The American Cancer Society (ACS) defines breast tumors as malignant (cancerous) or benign (benign) (Hamedani-KarAzmoudehFar et al., 2023).

Technology is growing swiftly in the sphere of medical diagnostics. The widespread adoption of computer-assisted diagnostics can be attributed to its exceptional outcomes in terms of precision and efficacy. Significantly involved is artificial intelligence, particularly convolutional neural network (CNN). Images of breast cancer may be further sorted into benign and malignant tumors using the software. The new technology can assist developing countries with a scarcity of professionals in the field of medical imaging and doctors and aid in early diagnosis (Aljuaid et al., 2022).

This study's major purpose is to compare several transfer learning strategies of pre-trained CNN for the categorization of benign and malignant tumors in mammography images. This process consists of the following steps:
1. Developing a method for automatically classifying breast cancer and benign masses from mammogram images.
2. To determine the optimal number of frozen layers in a pre-trained CNN.

The remainder of this paper is arranged as follows. Section 2 provides some related work. Section 3 presents the methodology by describing the scenarios of VGG-19. Experimental analysis and results are presented in Section 4, and Section 5 provides the conclusions.

## 2. Related work

Recent research has proposed several Computer-Assisted Diagnosis (CAD) systems for breast cancer detection. Zeiser et al. (2020) used U-net models for the DDSM dataset and obtained classification accuracy of 85.95%. Choudhary et al. (2021) Proposed a CNN model-based transfer learning strategy that is enhanced by structured filter pruning for breast cancer classification. Three pre-trained models, VGG19, ResNet34, and ResNet50 were utilized in their experiments. With the VGG19 pruned model, they achieved an accuracy of 91.25 %; with the ResNet34 pruned model, the accuracy rises to 91.80 %; and with the ResNet50 model, the classification accuracy rises to 92.07%.

Assari et al. (2022) have proposed transfer learning via GoogleNet for breast mass detection. The proposed model achieves a sensitivity of 90.91%. Mohapatra et al. (2022) evaluated the performance of various CNN architectures, such as AlexNet, VGG16, and ResNet50, by training some from scratch and others with pre-trained weights. The above-described classification models are trained and evaluated using mammogram images from the mini-DDSM dataset. Using rotation and zooming techniques increases the volume of data. The dataset has been divided into 90:10 portions. When trained with pre-trained weights, AlexNet demonstrated an accuracy of 65%, while VGG16 and ResNet50 demonstrated an accuracy of 65% and 61%, respectively. When trained from scratch, VGG16 performed significantly worse than AlexNet, while AlexNet outperformed others. When transfer learning was applied, VGG16 and ResNet50 performed well.

## 3. Methodology

In this study, five transfer learning scenarios will be evaluated based on their ability to detect breast cancer. Because VGG19(*Visual Geometry Group - University of Oxford*, n.d.) is a good CNN model that has been used successfully in a wide range of applications, it was chosen as a pre-trained model to study how freezing layers affect the accuracy of detection. Table 1 outlines these five scenarios in addition to the training of whole VGG19 layers. In the VGG19 model, the last fully connected layer (fc8) consists of one thousand neurons, each of which represents a single object. However, our study focuses on only two objects (benign or malignant), so the fc8 has been replaced with another layer that has only two neurons. The first is used to identify benign cells, while the second is used to identify malignant ones. All of the proposed scenarios are presented as shown in Figure 1. Where the blue layers represent the freezing blocks of convolution layers, the two green layers represent the fully connected layers (fc6 and fc7), the red layer represents the classifier layer with two objects (benign and malignant), and the final layer represents the Softmax layer.

**Table 1.** scenarios of transfer learning

| Scenarios | Description | Number of convolution layers | |
|---|---|---|---|
| | | freezing | training |
| One | Freeze first block | 2 | 14 |
| Two | Freeze first two blocks | 4 | 12 |
| Three | Freeze first three blocks | 8 | 8 |
| Four | Freeze first four blocks | 12 | 4 |
| Five | Freeze all blocks | 16 | 0 |
| Six | Train all blocks | 0 | 16 |

## 4. Experimental Analysis

### 4.1. Dataset

The BreaKHis database, including microbiology images of breast cancers, was utilized to analyze the six scenarios and select the best one (Spanhol et al., 2016). In this dataset, we only took a sample of 400x optical zoom. The dataset contains 547 benign and 1146 malignant samples. Figure 2 demonstrates these two types. This data was divided into 70% for training (383 benign, 802 malignant) and 30% for testing (164 benign, 344 malignant).

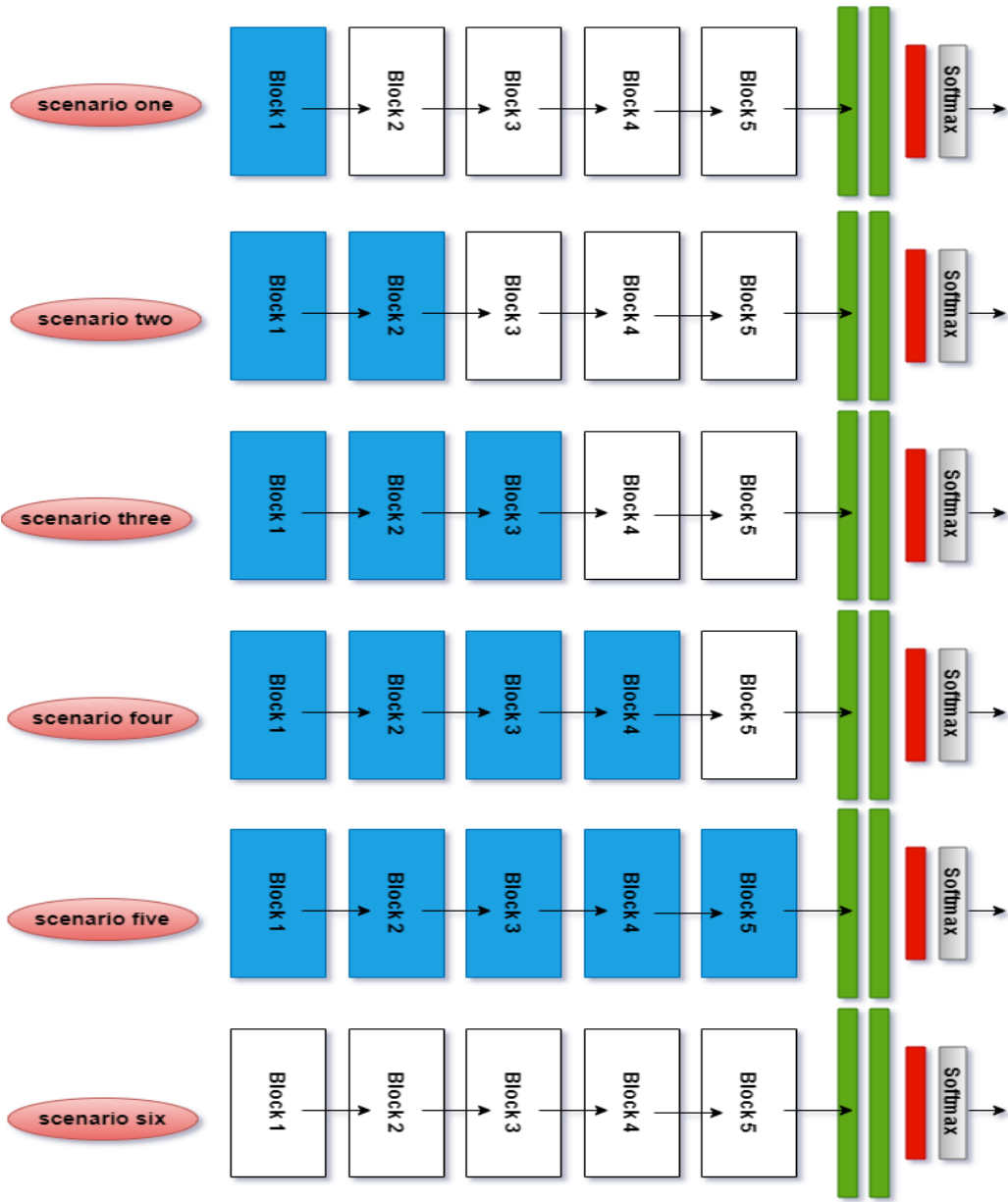

**Figure 1**: Research flow diagram

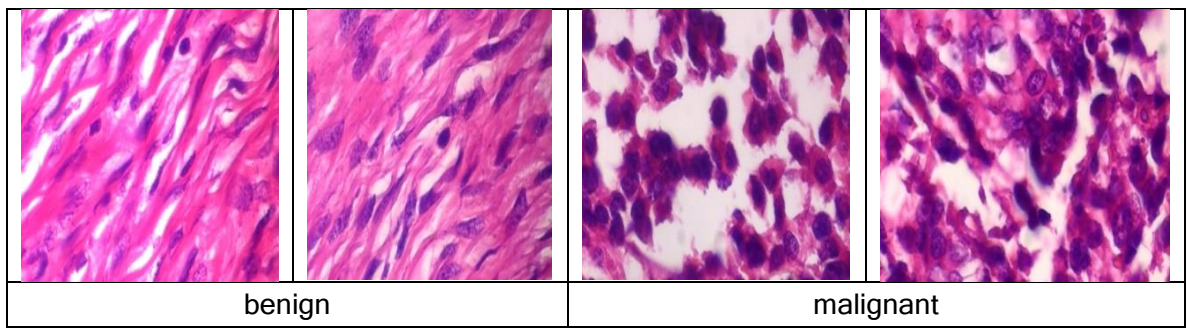

**Figure 2**: Various benign and malignant cases

## 4.2. Results and Discussion

In this section, the outcomes of the six different scenarios are presented. Each scenario in this experiment was trained using the same training set. Despite the fact that the majority of deep learning models are evaluated according to their overall accuracy, the quality of models in medical applications is associated with a high risk of both false positives and false negatives. Using more sophisticated diagnostic techniques, false-positive outcomes can be corrected. However, false-negative results are a significant concern, as breast cancer is fatal if left untreated. Consequently, sensitivity is a crucial factor in assessing any computerized medical application(Chatterjee et al., 2022).

Table 2 displays the results of each model. As can be seen, so if we reorder the scenarios descending based on the highest sensitivity achieved, we will get the first scenario ranked first, then the third scenario ranked second, then the sixth scenario ranked third, then the fourth scenario ranked fourth, followed by fifth scenario, and lastly, the second scenario.

**Table 2.** Benign and malignant recognition results.

| Scenarios | Description | Accuracy % | Sensitivity % | Specificity % |
|---|---|---|---|---|
| One | Freeze first block | 91.73 | **95.64** | 83.54 |
| Two | Freeze first two blocks | 88.98 | 92.44 | 81.71 |
| Three | Freeze first three blocks | 90.55 | 94.77 | 81.71 |
| Four | Freeze first four blocks | 89.76 | 93.90 | 81.10 |
| Five | Freeze all blocks | 87.99 | 93.02 | 77.44 |
| Six | Train all blocks | 93.11 | 94.48 | 90.24 |

As shown in Figure 3, additional analysis was conducted by computing the confusion matrix for each model. According to the results, the first scenario was able to correctly classify the greatest number of malignant cases with 329 cases, compared to all other scenarios, and it performed better in detecting benign cases than freezing scenarios (from the second scenario to the fifth scenario), correctly classifying 137 cases. However, the sixth scenario where the entire model was trained yielded the best benign detection over all scenarios, with 148 cases.

Based on the results of training all six scenarios, we can conclude that transfer learning is capable of producing high detection rates for breast cancer cases. The best way to apply transfer learning is to only freeze the first layer block and train the remaining layers.

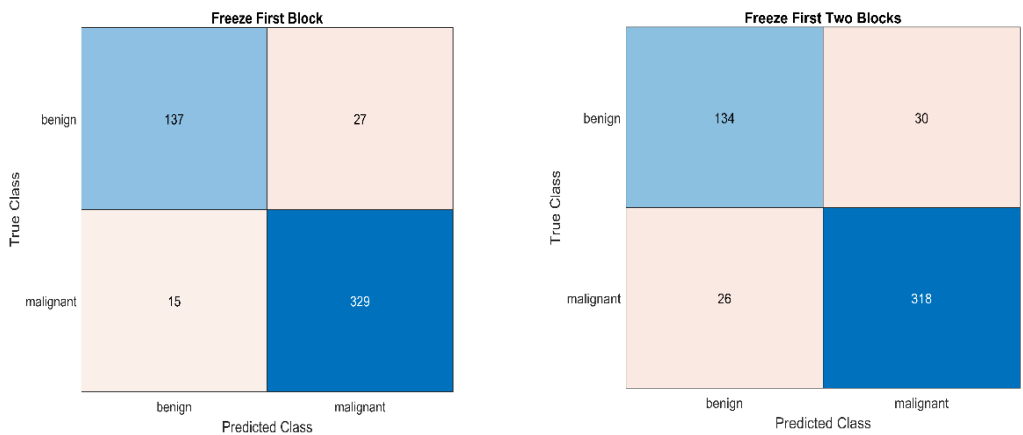

**Figure 3:** Confusion matrix for six scenarios.

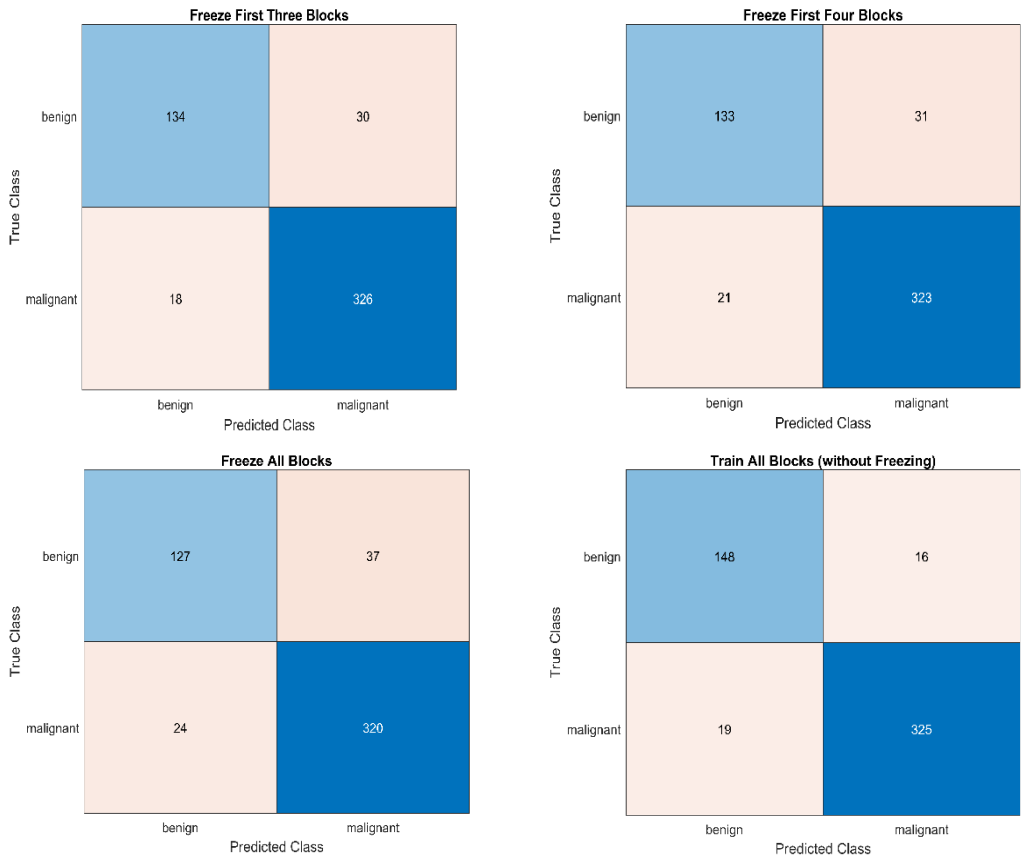

**Figure 3:** Confusion matrix for six scenarios. (continue)

## 5. Conclussion

This article discusses the effect of layers freezing in the transfer learning technique for working on breast cancer detection. The VGG19 model was used as a pre-trained CNN to conduct this study because it is efficient and simple. Six scenarios have been presented and divided based on the number of freezing blocks of layers in each scenario. Sensitivity was selected to evaluate these scenarios because of their vital impact on patient health. Based on the result, the freezing first block of VGG19 got the highest sensitivity, equaling 95.64%, while training the whole VGG19 got 94.48%. From these results, we can recommend freezing the first block only and training the last layers because the first block extracts the general features, while going deep in the layers will extract more abstracted features that will help increase classification accuracy.

As for future work, we will work on providing a comprehensive guide for researchers in the field of breast cancer detection by helping them select which pre-trained CNN model to use and how many freezing layers are required by studying a set of famous models like VGG16, GoogLeNet, ResNet50, ResNet101, and EfficientNet.

## Acknowledgments

This research is supported by the Malaysia Ministry of Higher Education (MOHE) Fundamental Research Grant Scheme (FRGS), no. FRGS/1/2019/ICT02/USM/03/3.

## References

Aljuaid, H., Alturki, N., Alsubaie, N., Cavallaro, L., & Liotta, A. (2022). Computer-aided diagnosis for breast cancer classification using deep neural networks and transfer learning. *Computer Methods and Programs in Biomedicine*, *223*, 106951. https://doi.org/10.1016/j.cmpb.2022.106951

Assari, Z., Mahloojifar, A., & Ahmadinejad, N. (2022). A bimodal BI-RADS-guided GoogLeNet-based CAD system for solid breast masses discrimination using transfer learning. *Computers in Biology and Medicine*, *142*, 105160.

https://doi.org/10.1016/j.compbiomed.2021.105160

Chatterjee, S., Biswas, S., Majee, A., Sen, S., Oliva, D., & Sarkar, R. (2022). Breast cancer detection from thermal images using a Grunwald-Letnikov-aided Dragonfly algorithm-based deep feature selection method. *Computers in Biology and Medicine*, *141*, 105027. https://doi.org/10.1016/j.compbiomed.2021.105027

Choudhary, T., Mishra, V., Goswami, A., & Sarangapani, J. (2021). A transfer learning with structured filter pruning approach for improved breast cancer classification on point-of-care devices. *Computers in Biology and Medicine*, *134*, 104432. https://doi.org/10.1016/j.compbiomed.2021.104432

Gardezi, S. J. S., Elazab, A., Lei, B., & Wang, T. (2019). Breast Cancer Detection and Diagnosis Using Mammographic Data: Systematic Review. *Journal of Medical Internet Research*, *21*(7), e14464. https://doi.org/10.2196/14464

Hamedani-KarAzmoudehFar, F., Tavakkoli-Moghaddam, R., Tajally, A. R., & Aria, S. S. (2023). Breast cancer classification by a new approach to assessing deep neural network-based uncertainty quantification methods. *Biomedical Signal Processing and Control*, *79*, 104057. https://doi.org/10.1016/j.bspc.2022.104057

Mohapatra, S., Muduly, S., Mohanty, S., Ravindra, J. V. R., & Mohanty, S. N. (2022). Evaluation of deep learning models for detecting breast cancer using histopathological mammograms Images. *Sustainable Operations and Computers*, *3*, 296–302. https://doi.org/10.1016/j.susoc.2022.06.001

Spanhol, F. A., Oliveira, L. S., Petitjean, C., & Heutte, L. (2016). A Dataset for Breast Cancer Histopathological Image Classification. *IEEE Transactions on Biomedical Engineering*, *63*(7), 1455–1462. https://doi.org/10.1109/TBME.2015.2496264

*Visual Geometry Group - University of Oxford*. (n.d.). Retrieved November 19, 2020, from https://www.robots.ox.ac.uk/~vgg/research/very_deep/

*WHO position paper on mammography screening*. (n.d.). World Health Organization. Retrieved August 19, 2022, from https://www.who.int/publications/i/item/who-position-paper-on-mammography-screening

Zeiser, F. A., da Costa, C. A., Zonta, T., Marques, N. M. C., Roehe, A. V., Moreno, M., & da Rosa Righi, R. (2020). Segmentation of Masses on Mammograms Using Data Augmentation and Deep Learning. *Journal of Digital Imaging*, *33*(4), 858–868. https://doi.org/10.1007/s10278-020-00330-4